**Title:** Estimated Incidence of Ophthalmic Conditions Associated with Optic Nerve Disease in Middle Tennessee

**Running title:** Incidence of Optic Nerve Disease in Middle TN

**Authors:** Shikha Chaganti, M.S., Katrina M. Nelson, B.S., Robert Harrigan, B.S., M.B.A., Kunal P. Nabar, Naresh Nandakumar, Tara Goecks, M.D., Seth A. Smith, Ph.D., Bennett A. Landman, Ph.D., Louise A. Mawn, M.D.

**Affiliations:**

S. Chaganti, B. A. Landman*, K. M. Nelson, R. L. Harrigan, K. Nabar, N. Nandakumar are with the Department of Electrical Engineering and Computer Science at Vanderbilt University, Nashville, TN, USA (correspondence e-mail: bennett.landman@vanderbilt.edu )

L. A. Mawn and T. Goecks are with the Vanderbilt Eye Institute, Vanderbilt University Medical Center, Nashville, TN, USA

S. Smith is with the Vanderbilt Eye Institute, Vanderbilt University Medical Center, Nashville, TN, USA

# ABSTRACT

*Aims*

The objective of this paper is to determine the incidence of ophthalmic disease potentially leading to optic nerve disease in Middle Tennessee.

*Methods*

We use a retrospective population-based incidence study design focusing on the population of middle Tennessee and its nearby suburbs (N=3 397 515). The electronic medical records for all patients evaluated or treated at a large tertiary care hospital and clinics with an initial diagnosis of a disease either affecting the optic nerve, or potentially associated with optic nerve disease, between 2007 and 2014 were retrieved and analyzed.

*Results*

18 291 patients (10 808 F) with 18 779 incidence events were identified with an age range of 0-101 years from the query of the Vanderbilt BioVU. Estimated age-adjusted incidence per 100 000 population per year was 198.4/145.1 (glaucoma F/M), 14.4/11.4 (intrinsic optic nerve disease F/M), 10.6/5.8 (optic nerve edema F/M), 6.1/6.6 (orbital inflammation F/M), and 23.7/6.7 (thyroid disease F/M). Glaucoma incidence was strongly correlated with age with the incidence sharply increasing after age 40. Optic nerve edema incidence peaked in the 25-34 old females. African American population has increased likelihood of glaucoma, orbital inflammation, and thyroid disease.

*Conclusions*

Mapping the incidence of pathologies of the optic nerve is essential to the understanding of the relative likelihood of these conditions and impacts upon public health. We find incidence of optic nerve diseases strongly varies by gender, age, and race which have not been previously studied using a unified framework or within a single metropolitan population

# **INTRODUCTION**

Incidence of optic nerve related ophthalmic disease is an important area for study as a host of conditions have the propensity to result in optic nerve damage, which often leads to irreversible vision loss and visual disability. Some of these processes are acute and may present with rapid visual deterioration (ischemic optic neuropathy, optic neuritis as examples), while others are more insidious and manifest over time (chronic open angle glaucoma). While these diseases may not initially present with optic nerve dysfunction in earlier stages on a clinical level, each disease process lends itself to compromise optic nerve integrity, resulting in optic nerve injury.

Mapping the incidence of pathologies of the optic nerve is essential to understanding the relative likelihood of these conditions and gives a framework upon which to target treatment. While several population-based studies have evaluated incidence and prevalence of various pathologic ophthalmic conditions one at a time[1-6], none have evaluated multiple disease processes in a diverse population base covering a large geographic area with rural and metropolitan components.

Electronic medical records (EMRs) offer the opportunity to glimpse at the health histories of millions of individuals without requiring centralized reporting of disease (i.e., registries). The focus of this paper is to characterize the incidence of five distinct ophthalmic conditions impacting the middle Tennessee area using the Vanderbilt University Medical Center (VUMC) anonymized (Synthetic Derivative) medical records. We evaluate the incidence of ophthalmic conditions affecting optic nerve function in five major disease cohorts: glaucoma[5, 7-9], intrinsic optic nerve disease[2, 6, 10-12], optic nerve edema (and idiopathic intracranial hypertension)[3, 4, 13-15], orbital inflammation[16-19], and thyroid disease[1, 20, 21]. The cohorts were investigated (as opposed to specific diagnoses) to ensure reliable patient-group assignment across thousands of patients without expert chart review of individual patients.

## METHODS

### *Study Population*

Nashville and its surrounding suburbs make up the most densely populated area in Tennessee. VUMC (in Nashville, TN) captures a large proportion of the patient base within the middle Tennessee corridor (including the Nashville metropolitan area), specifically zip codes with the three-digit prefixes 370, 371, 372, 373, 374, 383, 384, and 385. Based on 2010 US Census data[22], these areas have a combined population of 3 397 515; age by sex distributions were retrieved to facilitate more detailed investigation of incidence rates. Herein, we limit our focus to United States (US) zip codes for this middle TN catchment area. De-identified data from patients diagnosed with the diseases of interest were identified from the Vanderbilt Synthetic Derivative database (Fig. 1). This database contains the electronic medical records for over 2.2 million unique individuals. The search interface allows the user to input basic clinical and demographic information, such as International Classification of Diseases version 9 (ICD-9) codes, Current Procedural Terminology (CPT) procedure codes, medications, lab values, age, and gender. Note the census population estimate is for the median study year. From 2011 – 2013, 46% of patients admitted for an ophthalmic condition were admitted to VUMC as opposed to other area hospitals based on Tennessee state maintained inpatient hospitalization records of patients residing in the indicated zip codes (total events: 9 568), Prior hospitalization records with this level of detail were not available, and subsequent records had not been released as of the time of this study. Therefore, reference population estimates from the census data were scaled accordingly (e.g., market-adjusted population). The protocol, compliant with the Health Insurance Portability and Accountability Act (HIPAA), was approved by the Institutional Review Board of Vanderbilt University. The study adhered to the tenets of the Declaration of Helsinki.

### *Data Collection*

The ophthalmic conditions investigated in this study (glaucoma, intrinsic optic nerve disease, optic nerve edema, orbital inflammation, and thyroid disease) were defined by ICD-9 codes (Table 1). De-identified demographics (age, sex, ethnicity, and race) and all ICD-9 events (date / code pairs) were retrieved for all patients with ICD-9 codes matching Table 1 from the VUMC BioVU Synthetic Derivative (33 494 patient records with 4 286 224 ICD-9 events). The earliest records dated from 1994, but full adoption of the electronic medical record system did not take place until 2004. Therefore, 2007 was selected as the first year to minimize the misinterpretation of follow-up visits as initial diagnosis events. The BioVU Synthetic Derivative removes protected health information in the form of dates by time-shifting all events for each patient by a random (but fixed per patient) number of days. Patients retained for study met the following criteria: (1) had an office visit between the years 2007 and 2014, (2) had initial diagnosis of one of the ICD-9 codes of interest within the time range, (3) resided in the geographical region of interest (zip code prefix of 370, 371, 372, 373, 374, 383, 384, and 385), and (4) had complete age and gender information. The data from 2014 was not included in the final analysis as all of the data from 2014 was not available. For the purposes of this study, the date of occurrence of a condition was defined as the date of the first ICD-9 code with a cohort.

### *Data Analysis and Statistical Methods*

To ensure sufficient sample size, patients were binned according to 13 age groups represented in the US Census data (i.e., 0-4, 5-9, 10-14, 15-19, 20-24, 25-34, 35-44, 45-54, 55-59, 60-64, 65-74, 75-84, 85+ years of age). Note that the patient age is the age at time of initial diagnosis with an ICD-9 code within one of our five cohorts. A patient may belong to multiple groups if they had multiple conditions within the period of this study; note that the age of a patient could be different in different groups if the conditions occurred at different time points. The annual incidence rate was calculated by dividing the count of newly diagnosed patients within each time interval (e.g., total divided by 7 years for the complete study) by the estimated market-adjusted population (Fig. 1). Note that this approach averages both the incidence and health care system "market share" metrics over a 7-year period as to minimize the impact of any year-to-year variations (e.g., change in number of physicians, insurance contract negotiations, and population growth). A 95% confidence interval was constructed for each incidence estimate based on the Poisson distribution. Fisher's exact test was used to calculate the significance of relative likelihood of gender, age, and race incidence.

A co-occurrence graph for the 79 ICD-9 codes (Table 1), over the five cohorts was constructed. Each node represents an ICD-9 code and is color-coded by its respective cohort. Each edge in the graph represents a co-occurrence measure between two nodes called the *Dominant Confidence (DC)* measure using the ForceAtlas2 algorithm[23].

### *Validation*

A subset of 100 patients was selected with 20 subjects chosen uniformly at random from within each disease cohort. No duplicate subjects were selected across cohorts. Traditional expert chart review was performed to determine the frequency of coding errors to ascertain if the ICD-9 reflected the actual diagnostic state of the patients.

## RESULTS

### *Race, Gender, and Age Distribution of Patients*

Among the 3 397 515 residents of the defined area (market-adjusted population 1 321 255), a total of 18 779 diagnosis incidence events were found for 18 291 unique patients. This number corresponds to an age adjusted incidence of 168.1/100 000 residents (95% CI: 161.7-170.6).

The demographics of the patients in the cohort reflected the central Tennessee population with 1.5% Asian, 12.8% African American, <1% Native American, 80.8% Caucasian, and 4.5% other/multi-racial/unknown with 2% Hispanic, 88% non-Hispanic, and 9% unknown. Tables 2-3 present the patient counts by race, ethnicity, and age, while Supplementary Table 1 presents an analysis of cohort by year. We used Fisher's exact test by doing one-vs.-rest comparison, for the relative incidence of each race within a disease cohort, compared to the general population. It was observed that Caucasians were less likely to have incidence events in all of the cohorts. African Americans were 1.27 times more likely to have Glaucoma, 1.95 times more likely to have Orbital Inflammation, and 1.23 times more likely to have Thyroid Disease. Asians were 1.39 times more likely to have Glaucoma, and 1.81 times more likely to have Thyroid Disease, but significantly less likely to have Edema or intrinsic optic nerve disease than other races.

Incidence counts by gender controlled over age are shown in Supplementary Tables 2-6. The female to male ratio in the central TN population is almost equal (1.05). However, there was a clear gender bias in the patient population. Females were 1.2 times more likely to have Glaucoma than males ($p<0.05$). Female population was 2.1 times more likely to have optic nerve edema, 3.7 times more likely to have Thyroid Disease, and 1.3 times more likely to have intrinsic optic nerve disease (Supplementary Table 7).

Similar analysis of age (Supplementary Table 8) showed that Glaucoma is increasingly likely after the age of 55. Intrinsic optic nerve disease is more likely in people over 60. The likelihood of Thyroid disease increases with age as well, with a peak in 45-54 age group. Whereas, optic nerve edema is most likely in young women, peaking in the 25-34 age group. Orbital inflammation has the highest relative incidence in young children.

*Incidence Findings*

Incidence of each of the disease cohorts did not systematically differ across the time period of the study (e.g., lack of apparent trends in Supplementary Table 1). For the remainder of the analyses, the diagnoses are pooled across years and incidence is scaled to reflect the increased time period. Fig. 2 shows a co-occurrence graph for the 79 ICD-9 codes (Table 1) and table for the five cohorts.

*Validation*

For 94 of 100 patients, direct evidence of the reported ICD-9 code that led to cohort inclusion was found in the anonymized EMR (e.g., clinic notes, consult notes). Scanned handwritten notes and Adobe Portable Document Format (PDF) notes, which constitute a subset of the clinic and consult notes, are not included in the anonymized EMR due to the difficultly in removing protected health information at scale. Hence, the anonymized EMR is incomplete. For 5 such patients with missing notes, sufficient indirect evidence (including multiple entries of ICD-9 codes, medications, referrals for a consult, lab reports, etc.) was found such that a practicing ophthalmologist was convinced that the patient was legitimately diagnosed with the ICD-9 code in the chart. Finally, one patient had an entry of orbital cellulitis (ICD-9 376.01), but the chart showed periorbital cellulitis (ICD-9 373.13), which indicates a coding error sufficient so that the patient would have not met inclusion criteria for the orbital inflammation cohort. Therefore, 99 of 100 samples yielded correct cohort groups. Under a binomial model, the exact confidence interval on error rate is 0.02% to 5.45%, with a mean of 1%.

## DISCUSSION

This study provides insight into the incidence of multiple disease processes leading to optic nerve disease in a greater geographical area (with inclusion of patients from metropolitan areas and of various ethnicities) than would have been possible with a traditional chart review. Incidence studies of various eye diseases have been performed in large national health care systems such as in Beijing[24] and Sweden[25], and in a homogenous area of the United States, Olmstead county but have not been performed in a large metropolitan population in the United States. In the following sections, we discuss each cohort and the concordance of our incidence estimates with those reported in the literature.

### *Glaucoma*

Glaucoma was by far the greatest optic nerve disease burden in the greater TN population. Similar to other studies, the incidence of glaucoma in the middle TN population dramatically increases with age over 40[5]. We see this trend very clearly, with the steep increase of the relative incidence of Glaucoma after 50 in Supplemental Table 8. The prevalence of open angle glaucoma in the United States has been estimated to be 2.2 million people. Our data was confined to new diagnosis of all forms of glaucoma and may underestimate the disease burden from glaucoma in the middle TN population. Chronic open angle glaucoma is the most common cause of optic neuropathy in adults over the age of 40 and is estimated to affect 2.2 million Americans currently, with an estimated 3.36 million in 2020[5]. Race is a large factor in the risk of visual field loss and blindness from glaucoma. In a recent study, patients of African descent with the same access to care were shown to have a hazard ratio of up to 3.61 in the development of visual field loss at intraocular pressures greater than 22mm Hg, as compared to patients of European descent[26]. In our patient population, we found that glaucoma had only a slightly higher incidence in African American population (Odds Ratio=1.27).

### *Intrinsic Optic Nerve Disease*

Disease related to the optic nerve other than glaucoma were pooled into an intrinsic optic nerve disease category and included optic neuritis, ischemic optic neuropathy, arteritic optic neuropathy. Optic neuritis most often affects young, Caucasian females, with a mean age of 32[27] with incidence of approximately 5 per 100 000[2]. Anterior ischemic neuropathy, including arteritic and non-arteritic subtypes, is the most common cause of acute optic neuropathy in adults over the age of 50. Its incidence has been reported at 2-10 per 100 000[10]. Arteritic anterior ischemic optic neuropathy comprises much fewer of the cases of anterior ischemic optic neuropathy (5-10%) and occurs most frequently in patients over 70 years old, with an estimated incidence of 10-19.8 per 100 000 patients[11, 12] and a peak incidence in women aged 70-79[11] By looking more broadly at the phenotype of optic neuropathies, we can appreciate population level trends with a great degree of accuracy. Notably, there is a moderate increase in incidence of optic neuropathy in females over males (13.5 versus 10.4 per 100 000). For both males and females, the incidence of optic neuropathy is low <35 years old and begins to increase with age. The overall levels of incidence are generally in agreement with the literature when factoring the multiple conditions captured in this cohort.

*Optic Nerve Edema*

The middle TN population cohort of patients with optic nerve edema included patients with idiopathic intracranial hypertension (IIH) and papilledema. A study of the incidence of IIH from 1988 reported 75 patients from 2 states and a later study from 2004 reported 51 patients from Mississippi. The later report which was from a poster presentation suggests that the incidence is rising with the increase in rates of obesity, with about 22.5 new cases each year per 100 000 overweight women of childbearing age[3, 28]. The IIH incidence has been reported to be 0.9-2.36 per 100 000[3, 13-15], though much higher in women[3, 4, 15] with a mean age of onset at presentation of 29.4 and average body mass index of 39.8. With differing etiologies leading to optic nerve edema (including intracranial hypertension and papilledema), prior studies may have provided a conservative impression of the overall incidence. Here, we find age-adjusted incidences of 12.7 (F) and 6.0 (M) per 100 000. As with prior studies, optic nerve edema is substantively elevated in women of childbearing age (20-54), with 2.24 increased incidence for 25-34 versus the rest of the population (Supplementary Table 8). Incidence in males peaked at 10-14 years old (14.0 per 100 000).

*Orbital Inflammation*

The category orbital inflammation includes a variety of conditions either from an infectious or idiopathic etiology to include orbital cellulitis, orbital myositis, orbital periostitis, dacryoadenitis, and non-specific orbital inflammation (NSOI), among others. With the broad range of conditions that lead to orbital inflammation, systematic incidence studies of individual disease entities, or even pooled conditions as in this study, at a population level do not appear to have been conducted with some exception for example sarcoidosis in the veteran population[29]. An increased incidence of orbital infection has been associated with socioeconomic deprivation[30]. We find substantially higher incidences in children and African American population (Tables 10 and 11).

### *Thyroid Disease*

We note that thyroid eye disease is not precisely / completely captured by ICD-9 diagnostic practices. Herein, we included 242.00 "Toxic diffuse goiter without mention of thyrotoxic crisis or storm" in the definition. While 242.00 includes eye disease resulting from hyperthyroidism, it is not specific. If we limit our consideration of thyroid eye disease to ICD-9 codes of 376.2*, these patients have definitive eye involvement, yet, only a small fraction of the eye patients would be found. For example, the 1523:388 patients in Supplementary Table 6 would drop to 139:50. Hence, this study presents the incidence of thyroid disease as a risk factor / prodromal aspect of thyroid *eye* disease. In[31], the authors found that 12.5% of thyroid patients present with eye involvement. Therefore, it would be reasonable to divide the thyroid disease incidence estimates by eight to yield a conservative estimate of thyroid *eye* disease. The ambiguity in clinical disease identification of thyroid eye disease remains in ICD-10. Thyroid eye disease typically presents in the setting of hyperthyroidism. Approximately 1.3% of the United States population is estimated to have hyperthyroid disease[32]. Approximately 50% of hyperthyroid patients have Graves' disease and of those patients with Graves' disease, approximately 25% have clinically significant thyroid eye disease[33]. Optic nerve compression is estimated to occur in 3-5% of patients with thyroid eye disease[20, 21]. In our cohort, 84.21% of patients had a diagnosis of thyrotoxicosis without mention of thyrotoxic crisis or storm (ICD9 242.00) also had a diagnosis of endocrine exophthalmos (ICD9 376.2, 376.21, or 376.22). Using these US populations studies, our estimated calculation of patients with thyroid disease in this middle TN cohort is 34.7 per 100 000 population. A study performed by Bartley et al detailed the incidence of thyroid ophthalmopathy in a cohort of patients in Minnesota, yielding 16.0 cases per 100 000 population per year for women and 2.9 cases per 100 000 per year for men[1]. Compared with Olmsted County, middle Tennessee is more diverse and more urban; substantially more thyroid disease patients were identified in this study (1 523 in 7 years versus 519 in 15 years). The overall incidence seen in this study was approximately two-fold higher (versus 16.0 per 100 000 for females and 2.9 per 100 000 for males). We note that the Bartley study was based on manual chart review and excluded patients with hypothyroidism and Hashimoto's thyroiditis without ophthalmopathy, which could have resulted in a systematic bias between manual review and ICD-9 studies. Moreover, diagnostic practices and technologies have substantively evolved in the two decades between these two studies. In summary, the current study indicates a substantially greater incidence of thyroid disease in the general population than previous studies have indicated.

### *Limitations and Conclusion*

A central limitation of the study is that incidences are reflective of physician diagnostic standards and insurance coding practices, which are reasonable proxy for true disease occurrence, but not identical. Second, the population of the target areas is based on census data from a median year. We chose not to linearly interpolate between the census years (as in[12]) given the growth of the Nashville area and hesitancy to extrapolate population beyond the census period; rather we pool across years as to level out inconsistencies due to systematic factors (i.e., those that may lead to year-by-year variations in market share). The lack of trends in incidence by year (Supp. Table 4) lends support to the selected study design methodology. Finally, the market share corrected population is based on reported inpatient data, whereas the patient cohorts are based on both inpatient and outpatient information. Outpatient information is not reportable to the state and is not available for institutions outside of the VUMC affiliated hospital network. If all patients had come to VUMC (and none to other hospital systems), incidence could not be less than a third of the reported values (1/0.389). Hence, small errors in market share analyses translate to small errors in incidence estimates. Moreover, we find that coding errors are rare for our target population (approximately 1:100) and would not substantively impact the incidence estimates. Importantly, these limitations are generally cohort- and demographic agnostic (i.e., would not substantively impact the reporting of one condition relative to another or one patient versus another). Therefore, we have high confidence in the incidence comparisons within and across Supplemental Tables (e.g., males versus females, age groups, optic nerve disease versus orbital inflammation).

In conclusion, this manuscript has characterized the incidence of five important sets of pathological conditions that impact the optic nerve through use of anonymized EMRs. In contrast to several previous studies, we estimate the incidence of conditions that are known to lead to optic nerve damage as opposed to incidence of optic nerve damage concomitant with diagnosis of a condition. The incidences reported in this study are generally higher than could be understood from the literature when individual conditions were more narrowly defined, or when incidence studies were performed in less diverse settings. Moreover, the reported incidence rates are reflective of current diagnostic technologies and practices. Hence, this study serves as an important update on the incidence of conditions that impact the optic nerve with demographics reflective of the US population (with a slight under-sampling of Asian and Hispanic and oversampling of African American).

## ACKNOWLEDGEMENTS

Funding/Support: Supported in part by an unrestricted grant to the Vanderbilt Eye Institute and Physician Scientist Award from Research to Prevent Blindness, New York, NY. This project was supported by NIH 1R03EB012461 and the National Center for Research Resources, Grant UL1 RR024975-01 (now at the National Center for Advancing Translational Sciences, Grant 2 UL1 TR000445-06). The content is solely the responsibility of the authors and does not necessarily represent the official views of the NIH. This work was conducted in part using the resources of the Advanced Computing Center for Research and Education at Vanderbilt University, Nashville, TN. This research was supported by NSF CAREER 1452485 and NIH grants 5R21EY024036. This research was conducted with the support from Intramural

Research Program, National Institute on Aging, NIH. This project was supported in part by ViSE/VICTR. This work was also supported by the National Institutes of Health in part by the National Institute of Biomedical Imaging and Bioengineering training grant T32-EB021937. The sponsor or funding organization had no role in the design or conduct of this research.

**REFERENCES**

1. Bartley GB, Fatourechi V, Kadrmas EF, Jacobsen SJ, Ilstrup DM, Garrity JA *et al.* The Incidence of Graves' Ophthalmopathy in Olmsted County, Minnesota. *American Journal of Ophthalmology* 1995; **120**(4)**:** 511-517.

2. Wilhelm H, Schabet M. The Diagnosis and Treatment of Optic Neuritis. *Dtsch Arztebl Int* 2015; **112**(37)**:** 616-626.

3. Durcan P, Corbett JJ, Wall M. The incidence of pseudotumor cerebri: Population studies in iowa and louisiana. *Archives of Neurology* 1988; **45**(8)**:** 875-877.

4. Kesler A, Stolovic N, Bluednikov Y, Shohat T. The incidence of idiopathic intracranial hypertension in Israel from 2005 to 2007: results of a nationwide survey. *Eur J Neurol* 2014; **21**(8)**:** 1055-1059.

5.  Friedman DS, Wolfs RC, O'Colmain BJ, Klein BE, Taylor HR, West S *et al.* Prevalence of open-angle glaucoma among adults in the United States. *Arch Ophthalmol* 2004; **122**(4)**:** 532-538.

6.  Beri M, Klugman MR, Kohler JA, Hayreh SS. Anterior ischemic optic neuropathy. VII. Incidence of bilaterality and various influencing factors. *Ophthalmology* 1987; **94**(8)**:** 1020-1028.

7.  Quigley HA, Dunkelberger GR, Green WR. Chronic Human Glaucoma Causing Selectively Greater Loss of Large Optic Nerve Fibers. *Ophthalmology* 1988; **95**(3)**:** 357-363.

8.  Quigley HA, Green WR. The Histology of Human Glaucoma Cupping and Optic Nerve Damage: Clinicopathologic Correlation in 21 Eyes. *Ophthalmology* 1979; **86**(10)**:** 1803-1827.

9.  Quigley HA, Addicks EM, Green W. Optic nerve damage in human glaucoma: Iii. quantitative correlation of nerve fiber loss and visual field defect in glaucoma, ischemic neuropathy, papilledema, and toxic neuropathy. *Archives of Ophthalmology* 1982; **100**(1)**:** 135-146.

10. Katz DM, Trobe JD. Is there treatment for nonarteritic anterior ischemic optic neuropathy. *Curr Opin Ophthalmol* 2015; **26**(6)**:** 458-463.

11. Petri H, Nevitt A, Sarsour K, Napalkov P, Collinson N. Incidence of giant cell arteritis and characteristics of patients: data-driven analysis of comorbidities. *Arthritis Care Res (Hoboken)* 2015; **67**(3)**:** 390-395.

12. Chandran AK, Udayakumar PD, Crowson CS, Warrington KJ, Matteson EL. The incidence of giant cell arteritis in Olmsted County, Minnesota, over a 60-year period 1950-2009. *Scand J Rheumatol* 2015; **44**(3)**:** 215-218.

13. Wakerley BR, Tan MH, Ting EY. Idiopathic intracranial hypertension. *Cephalalgia* 2015; **35**(3)**:** 248-261.

14. Radhakrishnan K, Ahlskog J, Cross SA, Kurland LT, O'Fallon W. Idiopathic intracranial hypertension (pseudotumor cerebri): Descriptive epidemiology in rochester, minn, 1976 to 1990. *Archives of Neurology* 1993; **50**(1)**:** 78-80.

15. McCluskey G, Mulholland DA, McCarron P, McCarron MO. Idiopathic Intracranial Hypertension in the Northwest of Northern Ireland: Epidemiology and Clinical Management. *Neuroepidemiology* 2015; **45**(1)**:** 34-39.

16. Zenone T. Orbital myositis and Crohn's disease. *Int J Rheum Dis* 2014; **17**(4)**:** 481-482.

17. Vargason CW, Mawn LA. Orbital Myositis as Both a Presenting and Associated Extraintestinal Sign of Crohn's Disease. *Ophthal Plast Reconstr Surg* 2015.

18. Katsanos A, Asproudis I, Katsanos KH, Dastiridou AI, Aspiotis M, Tsianos EV. Orbital and optic nerve complications of inflammatory bowel disease. *J Crohns Colitis* 2013; **7**(9)**:** 683-693.

19. Grimson BS, Simons KB. Orbital inflammation, myositis, and systemic lupus erythematosus. *Arch Ophthalmol* 1983; **101**(5)**:** 736-738.

20. Barrett L, Glatt HJ, Burde RM, Gado MH. Optic nerve dysfunction in thyroid eye disease: CT. *Radiology* 1988; **167**(2)**:** 503-507.

21. Bartalena L. Sight-Threatening Graves' Orbitopathy. In: De Groot LJ, Beck-Peccoz P, Chrousos G *et al.* (eds). Endotext: South Dartmouth (MA); 2000.

22. Bureau USC. Population Estimates. Available at: https://www.census.gov/popest/data/.

23. Jacomy M, Venturini T, Heymann S, Bastian M. ForceAtlas2, a continuous graph layout algorithm for handy network visualization designed for the Gephi software. *PloS one* 2014; **9**(6)**:** e98679.

24. Group BRRDS. Incidence and epidemiological characteristics of rhegmatogenous retinal detachment in Beijing, China. *Ophthalmology* 2003; **110**(12)**:** 2413-2417.

25. Jin Y-P, de Pedro-Cuesta J, Söderström M, Stawiarz L, Link H. Incidence of optic neuritis in Stockholm, Sweden 1990–1995: I. Age, sex, birth and ethnic-group related patterns. *Journal of the neurological sciences* 1998; **159**(1)**:** 107-114.

26. Khachatryan N, Medeiros FA, Sharpsten L, Bowd C, Sample PA, Liebmann JM *et al.* The African Descent and Glaucoma Evaluation Study (ADAGES): predictors of visual field damage in glaucoma suspects. *Am J Ophthalmol* 2015; **159**(4)**:** 777-787.

27. Dworak DP, Nichols J. A review of optic neuropathies. *Disease-a-month : DM* 2014; **60**(6)**:** 276-281.

28. Garrett J, Corbett J, Braswell R, Santiago M. The increasing incidence of IIH. The effect of obesity on frequency of occurrence in Mississippi. *Ann Neurol* 2004; **56**(suppl 8)**:** S69.

29. Birnbaum AD, French DD, Mirsaeidi M, Wehrli S. Sarcoidosis in the National Veteran Population: Association of Ocular Inflammation and Mortality. *Ophthalmology* 2015; **122**(5)**:** 934-938.

30. Johnston NR, Sanderson G. Orbital infection in New Zealand: increased incidence due to socioeconomic deprivation and ethnicity. *The New Zealand Medical Journal (Online)* 2010; **123**(1320).

31. Lazarus JH. Epidemiology of Graves' orbitopathy (GO) and relationship with thyroid disease. *Best Practice & Research Clinical Endocrinology & Metabolism* 2012; **26**(3)**:** 273-279.

32. Hollowell JG, Staehling NW, Flanders WD, Hannon WH, Gunter EW, Spencer CA *et al.* Serum TSH, T4, and thyroid antibodies in the United States population (1988 to 1994): National Health and Nutrition Examination Survey (NHANES III). *The Journal of Clinical Endocrinology & Metabolism* 2002; **87**(2)**:** 489-499.
22

33. Lazarus JH. Thyroid hormones and cognitive function. *Expert Review of Endocrinology & Metabolism* 2012; **7**(4)**:** 365-367.

# TITLES AND LEGENDS TO FIGURES

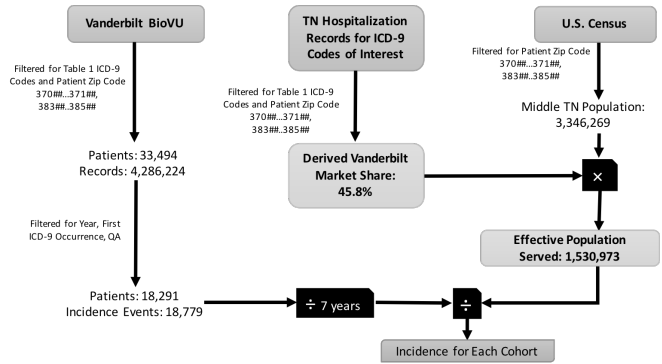

Fig. 1. Flowchart of calculation of average incidence per 100 000.

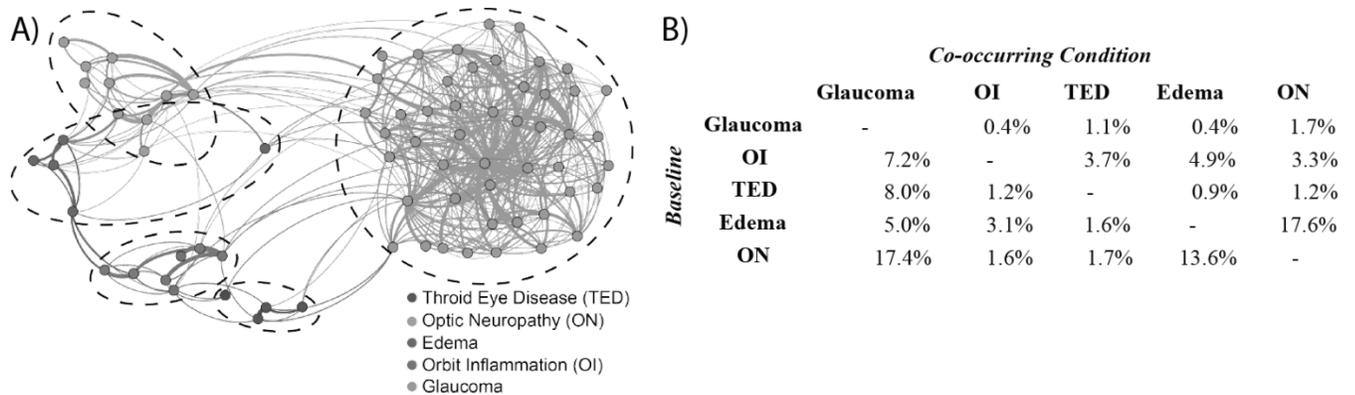

Fig. 2. ICD-9 codes within a cohort have a very high co-occurrence, however there is overlap between diagnoses cohorts as well, as quantitatively shown in the table at right. (A) shows a co-occurrence graph for the 79 ICD-9 codes (Table 1), over the five cohorts. Each node represents an ICD-9 code and is color-coded by its respective cohort. Each edge represents a co-occurrence measure between two nodes called the *Dominant Confidence (DC)* measure. The thickness of the edge represents the magnitude of *DC*, i.e. a thicker edge indicates a stronger co-occurrence. The *DC* measure between *A* and *B* is given by $DC(A,B) = max\left(\frac{P(A,B)}{P(A)}, \frac{P(A,B)}{P(B)}\right)$, where *P(A)* is the probability of having diagnosis *A*, and *P(A, B)* is the probability of having both diagnosis *A* and *B*. The graph was created by calculating the *DC* measure for all 6 241 pairs of ICD9 codes. The graph clustering was performed using ForceAtlas2 algorithm. The ForceAtlas2 algorithm computes attraction and repulsion forces proportional to the weights of the edges (*DC measures)* such that nodes with high co-occurrence are clustered closer together, while nodes with low co-occurrence are further away from each other. ICD-9 codes within a cohort have a very high co-occurrence, however there is overlap between diagnoses cohorts as well, as quantitatively shown in the table at right. In (B), each entry of the table shows the probability of a patient having the column condition given that they have the condition in the row.

Table 1. Cohort inclusion criteria

| Cohort | ICD-9 codes | Description |
| --- | --- | --- |
| Glaucoma | 365.0* | Borderline glaucoma |
| | 365.1* | Open-angle glaucoma |
| | 365.2* | Primary angle-closure glaucoma |
| | 365.3* | Corticosteroid-induced glaucoma |
| | 365.4* | Glaucoma associated with congenital anomalies, dystrophies, and systemic syndromes |
| | 365.5* | Glaucoma associated with disorders of the lens |
| | 365.6* | Glaucoma associated with other ocular disorders |
| | 365.7* | Glaucoma stage, unspecified |
| | 365.8* | Other specified forms of glaucoma |
| | 365.9* | Unspecified glaucoma |
| Intrinsic Optic Nerve Disease | 377.3* | Optic Neuritis |
| | 377.4* | Other disorders of optic nerve |
| Papilledema | 348.2 | Idiopathic intracranial hypertension |
| | 377.0, 377.00 | Papilledema |
| | 377.01 | Papilledema, increased intracranial pressure |
| | 377.02 | Papilledema, decreased ocular pressure |
| Orbital Inflammation | 376.0, 376.00 | Acute inflammation of orbit |
| | 376.01 | Orbital cellulitis |
| | 376.02 | Orbital periostitis |
| | 376.1 | Chronic inflammation of orbit |
| | 376.11 | Orbital granuloma |
| | 376.12 | Orbital myositis |
| | 373.13 | Abscess of eyelid |
| Thyroid Disease | 242.00 | Toxic diffuse goiter without thyrotoxic crisis or storm |
| | 376.2 | Endocrine exophthalmos |
| | 376.21 | Thyrotoxic exophthalmos |
| | 376.22 | Exophthalmic ophthalmoplegia |

TABLE II: Patient counts by race and ethnicity at VUMC

|  | Glc F | Glc M | ON F | ON M | Edema F | Edema M | OI F | OI M | TED F | TED M | Total |
|---|---|---|---|---|---|---|---|---|---|---|---|
| Asian |  |  |  |  |  |  |  |  |  |  | 361 |
| H. | 4 | 5 | 0 | 0 | 0 | 0 | 0 | 0 | 0 | 0 | 9 |
| NH | 153 | 121 | 7 | 3 | 2 | 2 | 2 | 4 | 42 | 11 | 347 |
| Unk | 2 | 2 | 1 | 0 | 0 | 0 | 0 | 0 | 0 | 0 | 5 |
| African American |  |  |  |  |  |  |  |  |  |  | 2 865 |
| H. | 2 | 3 | 2 | 1 | 0 | 0 | 0 | 0 | 1 | 0 | 9 |
| NH | 1 158 | 927 | 89 | 60 | 115 | 30 | 75 | 74 | 238 | 46 | 2812 |
| Unk | 13 | 16 | 0 | 3 | 1 | 1 | 1 | 1 | 6 | 2 | 44 |
| American Native |  |  |  |  |  |  |  |  |  |  | 33 |
| H. | 2 | 2 | 1 | 0 | 0 | 0 | 0 | 0 | 1 | 0 | 6 |
| NH | 13 | 11 | 0 | 0 | 0 | 0 | 1 | 0 | 2 | 0 | 27 |
| Unk | 0 | 0 | 0 | 0 | 0 | 0 | 0 | 0 | 0 | 0 | 0 |
| White | 0 | 0 | 0 | 0 | 0 | 0 | 0 | 0 | 0 | 0 | 13 357 |
| H. | 88 | 64 | 13 | 16 | 11 | 9 | 26 | 21 | 28 | 14 | 290 |
| NH | 5 370 | 4 121 | 544 | 399 | 514 | 253 | 203 | 208 | 1 044 | 275 | 12 931 |
| Unk | 54 | 46 | 8 | 4 | 7 | 2 | 3 | 3 | 8 | 1 | 136 |
| Unknown / Other |  |  |  |  |  |  |  |  |  |  | 2 163 |
| H. | 25 | 22 | 3 | 3 | 1 | 1 | 3 | 4 | 8 | 2 | 72 |
| NH | 234 | 182 | 14 | 8 | 9 | 5 | 3 | 9 | 20 | 9 | 493 |
| Unk | 694 | 545 | 67 | 56 | 46 | 17 | 5 | 15 | 125 | 28 | 1 598 |
| Total | 7 812 | 6 067 | 749 | 553 | 706 | 320 | 322 | 339 | 1 523 | 388 | 18 779 |

TABLE III: Patients Counts by Age Group

| Min Age | Glc | ON | Edema | OI | TED |
|---|---:|---:|---:|---:|---:|
| 0-4 | 111 | 100 | 42 | 171 | 14 |
| 5-9 | 103 | 55 | 68 | 64 | 22 |
| 10-14 | 164 | 44 | 102 | 65 | 40 |
| 15-19 | 164 | 69 | 113 | 34 | 101 |
| 20-24 | 156 | 52 | 83 | 25 | 89 |
| 25-34 | 513 | 147 | 223 | 54 | 300 |
| 35-44 | 917 | 176 | 163 | 73 | 376 |
| 45-54 | 1997 | 188 | 117 | 62 | 422 |
| 55-59 | 1476 | 85 | 34 | 29 | 154 |
| 60-64 | 1685 | 92 | 30 | 25 | 138 |
| 65-74 | 3592 | 152 | 39 | 40 | 170 |
| 75-84 | 2176 | 109 | 12 | 11 | 67 |
| 85+ | 825 | 33 | 0 | 8 | 18 |